\documentclass[10pt,twocolumn]{article}
\usepackage{ol2}

\begin{document}

\twocolumn[ 

\title{Non-paraxial Airy beams}


\author{Denis V. Novitsky$^{1}$ and Andrey V. Novitsky$^{2*}$}

\address{
$^1$ B.I. Stepanov Institute of Physics, National Academy of
Sciences of Belarus, Nezavisimosti~Avenue~68, 220072 Minsk, Belarus\\
$^2$ Department of Theoretical Physics, Belarusian State University,
Nezavisimosti~Avenue~4, 220030 Minsk, Belarus \\
$^*$Corresponding author: andrey.novitsky@tut.by}

\begin{abstract}
We report the propagation dynamics of Airy light beams under
non-paraxial conditions. It is studied using the general approach
which deals with Fourier expansion of the beam. We show the
transformation of the beam from the Airy form as the paraxiality
parameter decreases. The role of evanescent waves in non-paraxial
regime is discussed.
\end{abstract}


 ] 

During some last years it was reported \cite{Sivil07PRA} about
exciting experimental realization of the Airy light beams. There
were indeed confirmed the important non-diffracting and accelerating
properties of these beams. From the theoretical point of view the
Airy beams are described by the diffraction equation \cite{Sivil07}
\begin{equation}
\frac{\partial u}{\partial z} + \frac{1}{2 k} \frac{\partial^2
u}{\partial z^2} = 0,
\end{equation}
which is valid only for paraxial waves.

In the current Letter we study the properties of non-paraxial Airy
beams, which are the exact solutions of the Maxwell equations. We
will perform the theoretical analysis of propagation of non-paraxial
beams and this can offer the new problems for experimentalists in
the field.

Let us consider the light beam propagating along $z$-direction in
isotropic medium with dielectric permittivity $\varepsilon$ and
magnetic permeability $\mu$. We suppose that in the initial plane
$z=0$ the field is distributed according to the Airy function as
${\rm Ai}(x/x_0) \exp(a x/x_0)$, where $a$ is a decay factor
limiting the beam energy at $x<0$, $x_0$ is an arbitrary transverse
scale. Then this field changes during propagation according to the
Maxwell equations.

If $a=0$, we get to the ordinary Airy light beam, which is
non-diffractive as it was discovered in the pioneering work
\cite{Berry}. However, such a beam should possess the infinite
energy to be created and, therefore, cannot be realized
experimentally. Parameter $a$ confines the beam to be generated, so
that it is not more non-diffractive, but can be experimentally
generated. Note that there are other possibilities to confine the
beam \cite{Grossman}.

It is well-known that an arbitrary beam distribution in the plane
$z=0$ can be constructed using the plane waves. Each plane wave
propagates as it follows from the Maxwell equations:
\begin{eqnarray}
{\bf H} (x,z,k_x) = \left( c_1(k_x) {\bf e}_y - c_2(k_x) \frac{{\bf
e}_y \times {\bf k}}{k_0 \mu} \right) {\rm e}^{i k_x x + i k_z z},
\nonumber \\
{\bf E} (x,z,k_x) = \left( c_2(k_x) {\bf e}_y + c_1(k_x) \frac{{\bf
e}_y \times {\bf k}}{k_0 \varepsilon} \right) {\rm e}^{i k_x x + i
k_z z},
\end{eqnarray}
where the amplitudes $c_1$ and $c_2$ (in general, complex numbers)
describe polarization of partial plane waves. The dependencies on
the transverse wavenumber $k_x$ in the amplitudes $c_1$ and $c_2$
describe the profile of the beam. If $c_1 = 0$ ($c_2 = 0$) then wave
is TE (TM) polarized. Transverse $k_x$ and longitudinal $k_z$
wavenumbers are connected with the dispersion equation, so that $k_z
(k_x) = \sqrt{k_0^2 \varepsilon \mu - k_x^2}$, where $k_0 = \omega /
c$ is the wavenumber in vacuum. Also, the wavevector is of the form
${\bf k} = k_x {\bf e}_x + k_z {\bf e}_z$.

In further the dimensionless parameters are introduced: transverse
wavenumber $q = k_x x_0$, wavenumber in vacuum $\chi = k_0 x_0$,
transverse $\tilde{x} = x/x_0$ and longitudinal $\tilde{z} = z/k_0
x_0^2$ spatial ranges.

To obtain the Airy light beam in initial plane $z=0$ we assume that
the amplitudes are equal to \cite{Sivil07}
\begin{eqnarray}
c_{1,2} (q) = \frac{1}{2 \pi} A_{1,2} {\rm e}^{-a q^2} {\rm
e}^{i(q^3 - 3 a^2 q - i a^3)/3}, \label{amplc}
\end{eqnarray}
where $A_1$ and $A_2$ are complex constants to provide an arbitrary
beam polarization.

Now the Fourier transform (the superposition of the plane waves
incident at different angles) is for the propagation of {\it a
non-paraxial Airy beam}. For example, for the electric field of the
beam one reads
\begin{eqnarray}
&& {\bf E} (\tilde{x},\tilde{z}) = \frac{1}{2 \pi}
\int_{-\infty}^\infty \left( A_2 {\bf e}_y + \frac{A_1}{\varepsilon
\chi} (-q {\bf e}_z + \zeta(q) {\bf e}_x) \right) \nonumber \\
&& \times {\rm e}^{i q \tilde{x} + i \chi \zeta(q) \tilde{z}} {\rm
e}^{-a q^2} {\rm e}^{i(q^3 - 3 a^2 q - i a^3)/3} dq,
\label{Airi_solution}
\end{eqnarray}
where $\zeta(q) = \sqrt{\chi^2 \varepsilon \mu - q^2}$.

This beam is the superposition of the TE and TM Airy beams. Both of
them are described in the same manner. Therefore, we can investigate
only TE Airy beam. TM beam is characterized by another amplitude
coefficient $A$. TE Airy beam has only $y$ electric field component
and follows from Eq. (\ref{Airi_solution}), when $A_1 = 0$.

\begin{figure}[t!]
\includegraphics[scale=0.9, clip=]{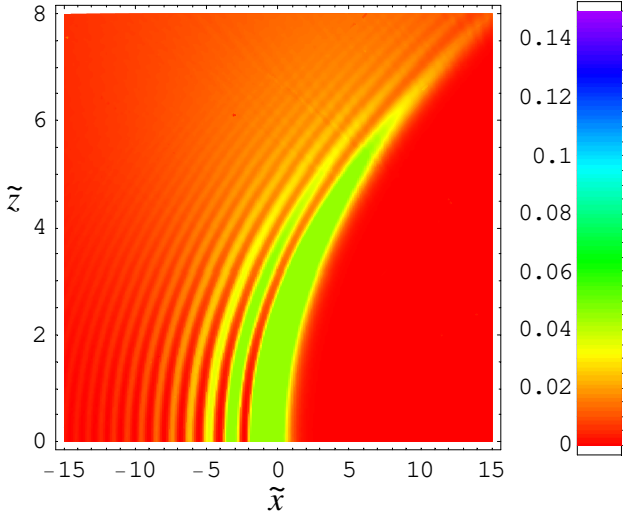}
\caption{The distribution of normalized intensity of the Airy beam
under paraxial condition ($\chi=100$) calculated using Eq.
(\ref{Airi_solution}). The decay parameter $a=0.1$. Refractive index
is $n=1$. }\label{fig:1}
\end{figure}

{\it Paraxial Airy beams}, which are usual for experiments
\cite{Sivil07, Besieris, Sivil08, Broky08}, can be obtained, when
the spectrum of propagating waves is wide. It corresponds to the
high-frequency radiation, which is described by the paraxiality
parameter $\chi>>1$. Then one can use the approximate expression
$\sqrt{\chi^2 \varepsilon \mu - q^2} \approx \chi \sqrt{\varepsilon
\mu} - \frac{q^2}{2 \chi \sqrt{\varepsilon \mu}}$, and the exact
solution (\ref{Airi_solution}) transforms to the paraxial one
\begin{eqnarray}
E_y (\tilde{x},\tilde{z}) = \frac{{\rm e}^{i n \chi^2 \tilde{z}}}{2
\pi} \int_{-\infty}^\infty  A_2 {\rm e}^{i q \tilde{x} - i \frac{q^2
\tilde{z}}{2 n}} \nonumber \\
\times {\rm e}^{-a q^2} {\rm e}^{i(q^3 - 3 a^2 q - i a^3)/3} dq,
\label{Airi_parax_gen}
\end{eqnarray}
where $n = \sqrt{\varepsilon \mu}$ is the refractive index. The
integral (\ref{Airi_parax_gen}) can be expressed using the
closed-form expressions \cite{Sivil07} as
\begin{eqnarray}
E_y(\tilde{x},\tilde{z})=A_2 {\rm e}^{i n \chi^2 \tilde{z}} {\rm Ai}
\left[ \tilde{x} - (\tilde{z}'/2)^2 + i a \tilde{z}' \right] \nonumber \\
\times {\rm exp} \left( a \tilde{x} - a \tilde{z}^{\prime 2}/2 - i
(\tilde{z}^{\prime 3}/12 - a^2 \tilde{z}'/2 - \tilde{x}
\tilde{z}'/2)\right), \label{paraxanal}
\end{eqnarray}
where $\tilde{z}'=\tilde{z}/n$. The equivalence of Airy beam
descriptions given by Eq. (\ref{Airi_solution}) for large parameter
$\chi$ and paraxial formulae (\ref{Airi_parax_gen}) and
(\ref{paraxanal}) can be directly verified. Fig.~\ref{fig:1} shows
the distribution of normalized beam intensity
$|E_y(\tilde{x},\tilde{z})|^2/|A_2|^2$ calculated from Eq.
(\ref{Airi_solution}) in paraxial case (for great number $\chi$).
One can see the typical structure of Airy beam and its main property
-- acceleration -- and compare them with previously reported results
obtained on the basis of Eq. (\ref{paraxanal}) \cite{Sivil07,
Besieris}.

\begin{figure}[t!]
\includegraphics[scale=0.5, clip=]{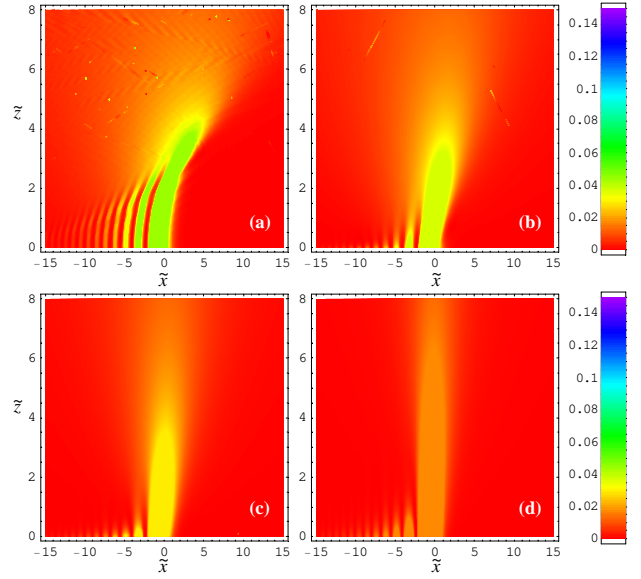}
\caption{The distribution of normalized intensity of the
non-paraxial Airy beam calculated from Eq. (\ref{Airi_solution}) at
(a) $\chi=5$, (b) $\chi=2$, (c) $\chi=1$, (d) $\chi=0.5$.
Parameters: $a=0.1$ and $n=1$.}\label{fig:2}
\end{figure}

As the paraxiality parameter $\chi$ decreases, the beam structure
becomes more and more different from that in Fig. \ref{fig:1}. When
coordinate $\tilde{z}$ is small, the non-paraxiality weakly
influences (see Fig. \ref{fig:2}(a)). The acceleration of the
non-paraxial Airy beam is the same as that of the paraxial one,
because it is determined by the amplitude spectrum (\ref{amplc}).
Only after $\tilde{z} = 2$ the beam starts to diverge rapidly. At
first, the side lobes disappear, while the main lobe is still
strong. After $\tilde{z} \sim 4$ the non-paraxial Airy beam
dissipates. The discussed Airy beam has the non-paraxiality
parameter $\chi$ comparable with the unity. The properties of the
paraxial beams quickly restore, if $\chi$ increases. For example,
for $\chi=10$ the spatial range, where the acceleration is
excellently seen, significantly raises (approximately by two times).
When $\chi$ further decreases, the beam's acceleration is getting
less and less pronounced (see Fig. \ref{fig:2} (b), (c), and (d)).
Now the beams propagate almost strongly ahead, while the side lobes
rapidly attenuate.

\begin{figure}
\centering \includegraphics[scale=0.85, clip=]{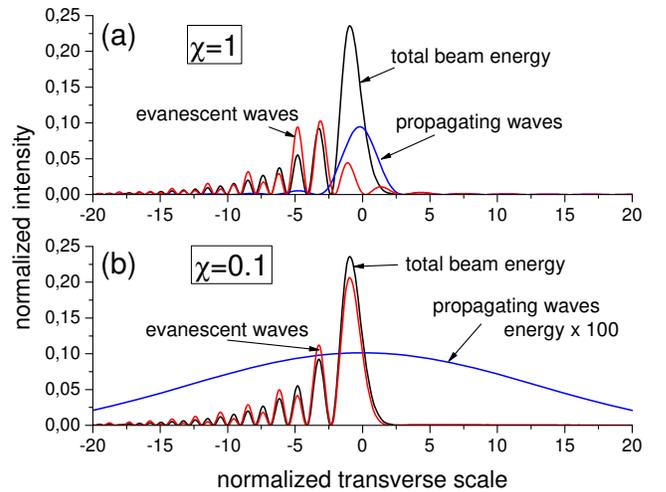}
\caption{The initial (at $\tilde{z}=0$) transverse distribution of
normalized intensity of the non-paraxial Airy beam at (a) $\chi=1$
and (b) $\chi=0.1$ carried by propagating waves, evanescent waves,
and both propagating and evanescent waves (total beam). Parameters:
$a=0.1$ and $n=1$.}\label{fig:3}
\end{figure}

The explanation of the behavior of the non-paraxial Airy beams can
be provided with dividing the spectrum of the plane waves forming
the beam into two parts. The first one describes {\it propagating
waves} and corresponds to the waves with real longitudinal
wavenumber $k_z$, therefore,
$-\chi\sqrt{\varepsilon\mu}<q<\chi\sqrt{\varepsilon\mu}$. The other
$q$s ($q<-\chi\sqrt{\varepsilon\mu}$ and
$q>-\chi\sqrt{\varepsilon\mu}$) are for the {\it evanescent waves},
which decay during propagation because of imaginary wavenumber
$k_z$. When parameter $\chi$ is great, the spectrum of the
propagating waves is wide, while evanescent waves can be neglected.
That is why one uses the paraxial approximation for the field of the
Airy beam (\ref{paraxanal}). When $\chi$ is comparable with the
unity, both propagating and evanescent waves should be taken into
account. In Fig. \ref{fig:3} we show the field distributions of
propagating and evanescent beams, when the total beam energy
corresponds to the Airy beam. For $\chi=10$ (not shown) the profile
of the propagating beam coincides with that of the total (both
propagating and evanescent) intensity distribution. The maximal
value in the profile of the evanescent waves is less than
$10^{-10}$, therefore, the evanescent waves can be certainly
neglected. For $\chi = 1$ the contributions of the evanescent and
propagating waves are comparable. The energy of the propagating
waves is concentrated in the main lobe, that is why it propagates at
long distance (see Fig. \ref{fig:2} (c)). The energy of the
evanescent waves is localized in the side lobes. This is the reason,
why the side lobes rapidly decay. In Fig. \ref{fig:3} (b) the beam
with very narrow spectrum of the propagating waves is shown. We can
observe very low energy of the propagating waves. This beam can be
approximately considered as the evanescent Airy beam. Note that the
sum of the energies of the propagating and evanescent waves is not
the total energy of the beam, because the last includes also the
interference term (it arises due to the interference of the
propagating and evanescent waves).

\begin{figure}[t!]
\includegraphics[scale=0.5, clip=]{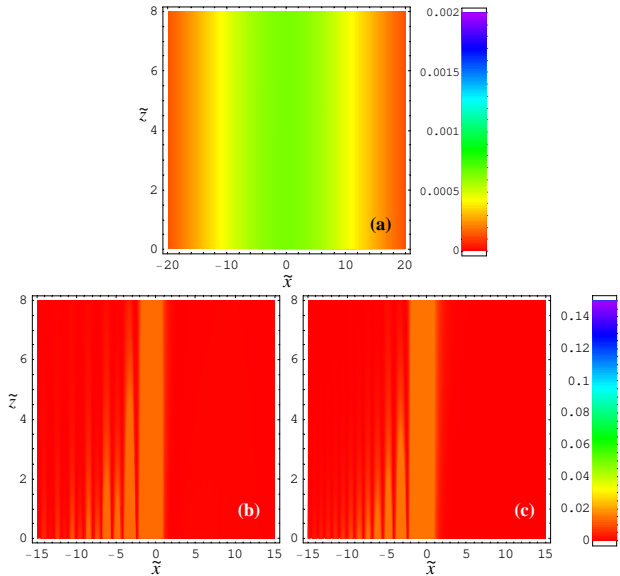}
\caption{The distribution of normalized intensity of the
non-paraxial Airy beam at $\chi=0.1$ corresponding to (a)
propagating waves, (b) evanescent waves, (c) total beam. Parameters:
$a=0.1$ and $n=1$. }\label{fig:4}
\end{figure}

In Fig. \ref{fig:4} the propagation of the almost evanescent Airy
beam is shown. The low intense propagating beam propagates without
spreading at the shown distance. However, it yields very low input
to the total intensity. In spite of the suppressed intensity of the
evanescent beam its main lobe propagates at long distance without
significant change of the intensity. It should be noted that the
realistic distance of beam's propagation decreases with diminution
of paraxiality parameter $\chi$, because $z=\chi x_0 \tilde{z}$.

As for the energy flux of the TE and TM Airy beams, it has $x$ and
$z$ non-zero components. More interesting situation arises for a
superposition of these partial beams, which realizes a specially
polarized Airy beam. Then the Poynting vector possesses also $y$
component. However, this case require separate investigation.
Similar study for the Bessel beams was earlier performed in Ref.
\cite{novitsky07}.

In conclusion, we have proposed the non-paraxial Airy beam as the
exact solution of the Maxwell equations. We have theoretically
analyzed the propagation of this beam and interpreted its properties
using those of propagating and evanescent waves. We have proved the
non-paraxiality parameter $\chi$ to be large enough (at least $\chi
>10$) to provide the paraxial beam and its properties
(acceleration). If $\chi$ is small, the evanescent waves prevail and
the beam rapidly decays. The beam with small $\chi$ can be regarded
as the evanescent Airy beam.

\end{document}